# Magnetic flux imaging in a 3D superconductor integrated circuit

T. Ren[1], A. Glatz[2], B. Jankó[1,3], J. L. Mallek[4], S. K. Tolpygo[4], V. K. Vlasko-Vlasov[1,5*]

[1] *University of Notre Dame, Department of Physics & Astronomy, Notre Dame, IN 46556 USA*
[2] *Nothern Illinois University, Department of Physics, De Kalb, IL 60115 USA*
[3] *University of Notre Dame, Department of Chemistry & Biochemistry, Notre Dame, IN 46556 USA*
[4] *Massachusetts Institute of Technology, Lincoln Laboratory, Lexington, MA 02421 USA*
[5] *Materials Science Division, Argonne National Laboratory, Lemont, Illinois 60439, USA*

**ABSTRACT**

We report on imaging magnetic flux distributions in a multilayered superconductor integrated circuit which emerge during magnetization and after field cooling of the circuit in the DC magnetic field. The obtained complicated field maps expose the flux propagation across the patterned superconducting ground planes sandwiching layers with Josephson junction-based logic cells, fine wire grid around the functional units, and multiple superconducting fill structures located in different inner layers. The observed intricate flux distributions are explained by specific patterns of Meissner screening currents and superconducting critical currents in different mutually interacting parts of the integrated circuit. Our results provide important insights into possible ways of improving the protection of superconductor integrated circuits from magnetic fields and their resilience against flux trapping.

# Introduction

Modern superconductor electronics (SCE) utilize multilayer designs containing Josephson junction (JJ) based circuits, resonators, filters, transmission lines, networks of wires connecting functional elements in various layers, and other components, typically sandwiched between two or more superconducting (SC) ground planes [1-4]. In most cases, the SC material of choice is niobium (Nb). The typical integrated circuit may contain up to ten superconducting thin-film layers patterned in various shapes and interconnected using superconducting vias. Millions of SC components in the circuits vary in dimensions from a deep sub-micrometer scale to tens and hundreds of micrometers. These components can produce extremely complicated, mutually intertwined current and magnetic flux distributions, especially in the presence of external magnetic fields, which can significantly impair the circuit operation. For instance, the appearance of Abrikosov vortices in superconducting layers near the JJs, or trapped magnetic flux quanta in multiply connected patterns, may introduce errors in logic operations or reduce the operation margins even in magnetically shielded SCE devices, and in extreme cases, can compromise circuit functionality entirely [5-7].

To reveal potential deficiencies and insufficiencies of the modern SCE design and fabrication, allowing the generation of undesirable vortices and magnetic flux trapping, we mapped magnetic induction patterns in an eight-layer superconductor integrated circuit [8], fabricated at MIT LL in the SFQ5ee fabrication process [9]. The magneto-optical imaging (MOI) technique [10] was used to visualize the induction distributions, $B(r)$, above the top surface of the circuit in the increasing and decreasing magnetic fields. We also imaged the $B(r)$ maps emerging after cooling the sample in a constant field. Remarkably, it was possible to see the flux patterns generated by structures in the top superconducting layer and by the SC components in the lower layers of the circuit. We used the critical state model to analyze the observed flux patterns. In this model, the penetration of Abrikosov vortices in separate SC film elements is followed by the flux filling of multiple apertures in their structure, and the flux front propagates consistently from the edges towards the middle of the entire circuit.

Since we are aware of magnetic imaging only in the assembly of rectangular SC strips placed in neighboring planes [11-12], we believe that the present work is the first demonstration of elaborate magnetic flux distributions in complex 3D superconducting structures.

## 1. Experiment

*a) Circuit design*

The circuit studied in this work is a multi-bit ac-driven shift register [8] with SC elements located in eight layers, M0-M7, as described in more detail in [5]. The circuit consists of units (bit-cells) with four resistively shunted Josephson junctions, four inductors, and a coupling transformer delivering an ac bias current driving operation of the shift register bit-cell. The units of 14 µm x 20 µm size are sandwiched between the bottom ground plane in layer M4, and the top ground plane in layer M7. They are connected in series along the y-direction and have 1-µm separation from their x-neighbors, forming a long meandering path containing 4513 cells in each of the six subcircuits of a 5 mm x 5 mm chip; see the optical picture in Fig. 1(a). The ground planes of the adjacent rows of cells are periodically connected by 4 μm bridges (V-bridges). As a result, the cumulative ground planes of each shift register become 480 μm x 2800 μm SC Nb film rectangles perforated by periodic 1-μm-wide slits (moats) with y-pitch of 40 μm or 100 μm, and x-pitch of 15 μm; see Fig. 1(b)-(c). The register input/output (I/O) circuitry and the common ac bias transformer primary are connected to the periphery contact pads which are located in all eight Nb layers. The pads are formed by continuous film rectangles, 0.5 mm long and 0.24 mm wide, in the top ground plane M7 and layers M2 and M0. In other layers, M1 and M3-M6, they are assembled of crossing strips, all connected by superconducting vias; see Sec. 2(d) for details.

Most elements of the circuit layers are made of 200 nm Nb film, except for the junctions' bottom electrode layer, M5, which is 135 nm thick. Nb layers are separated with planarized 200 nm thick $SiO_2$ dielectric spacers, except for the spacer between circuit layers M5 and M6, which is 280 nm thick. The Josephson junctions in the bit-cells are Nb/Al-AlO$_x$/Nb tunnel structures; see Fig. 6 in [1] and Fig. 2 in [9] for more details. The active circuit is surrounded by a grid of 2-μm-wide wires (W-grid in Fig.1(c)) on a 20-μm pitch, formed on the ground plane layer M4. This grid is connected to the two ground planes of the circuit in multiple points. It is designed to screen a residual magnetic field and provide distributed return paths for currents between the ground planes and the periphery contact pads.

In Fig.1(a), the top ground planes of six subcircuits with different periods of slits are visible as bright wide strips and are referred hereafter as GP-strips. They are separated by vertical sections of the W-grid; see Fig. 1(c). The slits in the GP-strips, periodically interrupted by V-bridges, form vertical slit-lines, which intermittently end at the bottom edge of the GP-strip or at some distance from the edge. Due to the arrangement of the bit-cells, this order changes at the top GP-strip edge, forming the above-mentioned meandering path, as shown by curved arrows in Fig. 1(b).

A fill structure of small 6 μm x 6 μm Nb film squares (S-squares in Fig. 1(c)) is placed in all layers outside the registers, and in the layers M0-M3 also under the registers (Fig. 1(d)). These fill structures assist in planarization of the circuit layers by chemical mechanical polishing. The projections of the S-squares partially cover some of the slits in the GP-strips while other slits remain uncovered (C-slits and U-slits in Fig. 1(c)) in the periodic fashion. We refer to the vertical lines of appropriate slits as C-lines and U-lines.

### b) Imaging magnetic flux

To study magnetic flux distributions, we used the MOI technique based on garnet indicator films with large Verdet constant, developed at Argonne National Lab [10]. When such a film is placed on the surface of a sample carrying an inhomogeneous normal magnetic field ($B_z$), the local magnetic moments in the film get tilted from the plane by an angle proportional to the local value of $B_z$. This changes the local image intensity (I) observed in linearly polarized light due to the Faraday effect and, after I-$B_z$ calibration, gives the $B_z(r)$ map of the sample.

In our experiments, the 5 mm x 5 mm Si chip with the superconductor integrated circuit was mounted on the cold finger of an optical Montana cryostat, covered by a garnet indicator, cooled below the SC transition temperature (T< $T_c$=9.2 K), and imaged in polarized light microscope using a low-noise digital camera. The magnetic field perpendicular to the chip surface was applied using a cylindrical magnetic coil controlled by a custom-developed software which scheduled ramping the coil current in small steps and taking successive pictures of the flux distributions. The flux patterns were imaged in increasing and decreasing fields applied to the zero-field cooled (ZFC) sample and after cooling in a constant applied field (FC). Below we describe the peculiar $B_z$-maps obtained after the calibration of the MO images of the sample.

## 2. Results and discussion

*a) Magnetization of the circuit in perpendicular magnetic fields: General picture*

Figures 2(a)-(f) show the flux patterns emerging over the major area of the circuit in a large range of perpendicular magnetic fields applied at $T$=5 K after ZFC. Magnetic flux starts penetrating at the periphery and gradually moves deeper towards the middle of the sample. Initially, a noticeable concentration of $B_z$ occurs around the contact pads which is visible as bright stripes around the pad rectangles at the bottom in Fig. 2(a). This $B_z$ concentration is caused by the expulsion of the magnetic flux from the pads by the induced Meissner screening currents. It is typical for perpendicularly magnetized flat SC samples [13, 14, 21]. For a single SC rectangle, the field decreases strongly away from the superconductor edge, as $B_z(r) \sim |r|/(r^2-r^2_{edge})^{1/2}$ [13, 14]. However, in our case, $B_z$ around the pads is nearly constant in the gap between the pad and the nearest lines of the W-grid due to the screening action of the SC grid wires.

In turn, the screening by the W-grid results in the appearance of bright horizontal lines of enhanced $B_z$ between the contact pads and GP-strips; see the bottom part of Fig. 2(a). The average flux density decays in the y-direction, perpendicular to the edge of the chip, indicating that the superconducting W-grid provides partial shielding of the external field, as expected in the circuit design. Fig. 2(a) shows that, already in small fields, the flux penetrating through the W-grid reaches the registers and distinct modulations of $B_z$ appear around the slits in the GP-strips. These modulations are different in the registers differing by the length of the moats. The strongest $B_z$ variations are in the ground planes of register #3 with the longest moats. Here, in the GP-strip #3, we see the rows of bright and dark vertical streaks corresponding to regions of enhanced and decreased magnetic fields in the moats. The streaks appear around the positions of V-bridges between the slits, and their intensity decreases towards the center of the chip.

With ramping up the applied field, the contrast of the above discussed flux inhomogeneities increases and new features emerge; see Fig. 2(b). From the regions of the enhanced $B_z$ near the corners of the contact pads, light strips of flux propagate along the diagonals of the W-grid. Here the channels are formed delivering the magnetic flux towards the edges of the GP-strips and concentrating the field there. The flux starts penetrating along the slits and in the gaps between the GP-strips. The bright vertical lines of enhanced $B_z$ appear abruptly along separate slit-lines (see GP-strip #2 in Fig. 2(b)), indicating an instability in the flux dynamics, likely triggered by small edge imperfections at the slit-line entry. Similar sharp flux changes assisted by the thermo-magnetic jumps [15,16] are often observed in uniform and patterned SC films. A possible self-organized nature of the instability will be discussed later.

With further increasing $H_z$, additional flux channels appear across the W-grid between the contact pads and GP-strips; see Fig. 2(c). Also, more bright streaks emerge along the slit-lines in the GP-strips as well. Apart from these details, the average flux front advances up inside the ground planes (see the envelope of filled bright slits in the GP-strips 1-2 and 4-6 in Fig. 2(c)).

The flux penetration gradually progresses until all the GP-strips and the gaps between them are filled with flux (Fig. 2(d)). At this stage, the *average* flux distribution in the GP-strips becomes similar to the x-shaped critical state pattern in a homogeneous strongly anisotropic SC rectangle with a weak transverse critical current; see sketches in Fig.2(g-h).

With a subsequent decrease of $H_z$, the flux exits from the GP-strips towards their ends, mostly along the slit-lines (Fig. 2(e)). $B_z$ also decreases in the gaps between the strips and in the W-grid area. The fronts of the average flux exit from the GP-strips again resemble those in an anisotropic SC rectangle.

Finally, when $H_z$ is switched off (Fig. 2(f)), there remains an intricate pattern of the trapped vortex regions, which reveals itself as small bright spots of enhanced $B_z$ around the slits in the GP-strips and over the area of contact pads. It is accompanied by the negative $B_z$ features (dark contrast) around the regions of trapped flux.

Leaving aside the peculiar emerging $B_z$ patterns, the *general scenario* corresponding to the observed magnetization of the circuit is a successive entry of magnetic flux from the periphery towards the center of the superconducting "Swiss cheese slice" with multiple holes of various shapes and sizes. Importantly, the circuit is surrounded by the continuous wire grid which prevents direct access of the field to the sample interior. Below we analyze specific local details of the flux screening and flux penetration in different elements composing the 3D structure of the circuit, observed near the end of one of the GP-strips.

### *b) High-magnification MO images of flux distributions in GP-strips*

The main features of the flux patterns emerging in the sample are clearly visible over the GP-strip #3 containing the longest slits. The optical pictures of the studied region are shown in Figs. 1(b)-(c) and the corresponding 8-layer design is sketched in Fig. 1(d).

Figure 3 shows the $B_z$-maps emerging around the GP-strip #3 in the increasing (panels (a-d)) and decreasing (panels (e-f) fields. Pictures at the bottom of the figure, (ai-fi), present the extended view of the panels (a-f) in the region outlined by the dashed line in (a). As mentioned above, already at small applied $H_z$, there is a strong concentration of the flux around the contact pads limited by the nearest W-grid cells, which is visible as bright rectangular bands at the bottom of Fig. 3(a)). The screening currents enhancing the field here, are sketched in Fig.4(a). In turn, the sketch in Fig. 4(b) presents currents in the horizontal segments of the W-grid responsible for the periodic light lines of enhanced $B_z$ between the pads and the GP-strips in Fig. 3(a). A closer look reveals that these lines have brighter contrast on their side facing the sample edge and darker on their sides facing the interior. The small isolated S-squares inside the W-cells show a smaller screening effect, and the flux easily penetrates between them. In the vertical segments of the W-grid, participating in screening adjacent left and right cells, the currents are cancelled and do not induce additional fields. However, the horizontal wire segments carry a dominating unidirectional current component, shown in Fig. 4(b) by longer arrows. This screening current flows along the x-line of cells and impedes the flux propagation from the sample periphery towards the center. As a result, $B_z$ is enhanced on the sides of the x-wire segments facing the contact pads (brighter edge of grid wires in Fig. 3(a)) and is suppressed on their opposite sides facing the sample interior (darker edge of grid wires).

The remarkable flux penetration features appear in small fields at the slit-lines inside the GP-strip. Below the V-bridges separating individual slits, $B_z$ is noticeably enhanced (see bright regions marked by Up arrows in Fig. 3(a)). Simultaneously, regions of *negative* $B_z$ (opposite to the applied $H_z$) form above the V-bridges (dark contrast marked by Down arrow in Fig.3(a)). In Fig.4(c-d) we show a sketch of the bending SC currents responsible for the emerging sign-changing $B_z$ pattern. Note that the longer and stronger $B_z$-contrast appears along the U-slits which extend to the edge of the GP-strip, marked by a longer arrow in Fig.3(a). U-lines starting from these slits get filled with flux much faster in larger field.

Figure 3(b) illustrates the emergence of flux channels between the contact pads and the GP-strips in the increasing $H_z$. The channels of flux penetration ("Ch" in Fig. 3(b)) with a width of about three W-grid cells start near the corners of the contact pads and go at $45^0$ towards the GP-strips. Formation and directionality of these channels can be explained by current crowding [17, 18] and the resulting enhanced nonlinear flux diffusion [19, 20] in the sharp turns of the W-grid. The corresponding crowding points are marked by red dots in Fig. 4(a-b).

When magnetic flux enters a W-cell, e.g. shown yellow in Fig. 4 (b), new crowding points form in its corners, keeping the diagonal flux propagation along the wire grid which forms the $45^0$ flux channels. Inside the emerging channels, the screening effect of the S-squares becomes noticeable. Here $B_z$ remains small above the S-square projections (see darker spots inside grid cells of the flux channels in Fig. 3(b)), but increases around them. In turn, the SC currents redistribute in the wire grid around the flux channel, which is revealed by the enhanced $B_z$ at the *vertical* y-segments of the nearby W-cells in Fig. 3(b).

After the flux is delivered through the channels to the edge of the GP-strip, it starts filling the first row of U-slits marked by a solid horizontal arrow in Fig. 3(b). The process shows an instability resulting in an abrupt filling of a long line of U-slits marked by dashed arrow in Fig. 3(b). Similar instability causes random filling of different U-lines in larger $H_z$ (Fig. 3(c)), while the flux entry remains delayed in the C-slits. In the filled U-lines, the flux is concentrated around the V-bridges forming a bead-like pattern. The preferential flux filling of the U-lines is caused by the increased current density in their V-bridges, which starts from the first row of U-slits; see a sketch in Fig. 4(c). When the critical current density is reached in the V-bridge, Abrikosov vortices carrying the flux quanta, move across it and transfer the flux to the adjacent U-slit. After filling each new U-slit, the current is enhanced in the next V-bridge, resulting in a chain-like flux propagation along the U-line. Such a preferential flux filling is observed during the entire magnetization cycle.

During filling of the U-lines, the sharp current turns around the slit ends in the V-bridges become an efficient source of Abrikosov vortices. Vortices are generated there due to the current crowding and nonlinear flux diffusion effects similar to those in the corners of the W-grid. However, now the propagation of vortices into the neighboring GP-strip regions is limited by pinning. As a result, the localized high-vortex-density spots are formed around the bridges along the U-lines; see Fig. 3(b-c). At a larger magnification (see inset in Fig. 3(b)) these spots resemble a combination of four flux petals characteristic for SC samples with sharp inward corners.

Even in large $H_z$, when the flux penetrates the entire sample, $B_z$ remains stronger in U-lines than in the neighboring C-lines (Fig. 3(d)). At the same time, in large fields, shielding by the superconducting S-squares yields an inhomogeneous flux distribution in the areas between the C- and U-lines. We identify their location by overlapping the projections of the S-squares on the MO images, as shown in Fig. 3(bi). In large fields, the saw-tooth-like patterns of enhanced $B_z$, marked by arrow in Fig. 3(di), extend from bright U-slits between the S-squares. Also, periodic spots of increased $B_z$ appear between the S-squares along the C-lines. Simultaneously, four-petal flux clusters extend around V-bridges in U-lines (compare Figs. 3(ci) and (di)). The described $B_z$ inhomogeneities form in the body of the GP-strip by Abrikosov vortices, which are generated at the edges of the slits and move into areas of strong local fields modulated by the S-squares in lower layers. Qualitatively the same picture is maintained up to the maximum applied field of 200 Oe.

With subsequent decrease of the field, the flux exits primarily along the U-lines causing the reduction of contrast along the U-slits, while brighter spots of trapped flux remain around the slits, between the projections of S-squares; see Fig. 3(e). In the centers of the V-bridges in U-lines, $B_z$ also decreases and the image intensity drops, but the bright four-petal trapped flux pattern remains around the V-bridges; see Fig. 3(ei). In turn, periodic spots of enhanced $B_z$ remain between the S-squares along the C-slits. The described trapped flux regions around the C- and U-lines stay in the body of the GP-strip, although in the lower layers, the flux density between the S-squares decreases with ramping $H_z$ down.

Finally, when the external field is switched off (Fig. 3(f)), bright spots of the trapped flux remain between the S-square projections mainly around the slits in the C-lines. Along the U-lines and in their V-bridges the contrast drops essentially, and the intensity calibration reveals the inversion of $B_z$ there (see next section). In the area of the W-grid below the GP-strips, the flux density is strongly weakened. The trapped flux remains relatively strong only in the contact pads, while $B_z$ inverts in the band around them (marked by arrow N in Fig. 3(f)). The details of the contact pad magnetization are discussed in section *d*).

At higher temperatures, the magnetic patterns in both increasing and decreasing $H_z$ are qualitatively similar to those described above. Although they emerge at smaller fields due to reduced critical currents in all niobium components at larger T.

## *c) Induction profiles measured across the ground plane slits*

Figure 5 presents quantitative estimates of the flux inhomogeneities appearing in the circuit. It shows induction profiles measured across the slits in the GP-strip, $B_z(x)$, and $B_z(y)$ scans across the contact pad and the W-grid at 7.5 K. The field enhancement in the slits turns out to be significant even at small $H_z$, and becomes extremely strong when the flux fills the slits; see left $B_z(x)$ plots in Fig. 5(a)-(b). With further increasing field, the relative amplification factor, $B_z/H_z$, decreases, although the absolute field change $\Delta B_z$ across the slits remains very large; see Fig. 5(c)-(d). The corresponding changes of the maximum $B_z$ in the first row of U-slits, measured below the V-bridge, are shown in Fig. 6. With decreasing field (Fig. 5(e)), the $B_z(x)$ maxima, reflecting flux concentration in the U-slits, turn into minima corresponding to the advanced flux exit along them. After switching $H_z$ off (Fig. 5(f)), $B_z$ in U-slits becomes negative due to the fields of currents supporting trapped flux in the GP-strip and in the S-squares surrounding the C-slits.

The right plots in Fig. 5 show $B_z(y)$ scans taken across the contact pad and the major region of the W-grid surrounding GP-strips (see the vertical scan line in left panels of Fig.5). They confirm a considerable field enhancement in the band around the contact pad, from $B_z/H_z\sim 7$ at small fields, to ~2 in large $H_z$. In small fields, there is a noticeable flux concentration in front of the W-grid wires (on their side facing the contact pads) and a drop towards negative $B_z$ behind them (on their side facing the middle of the sample), which gradually decreases towards the GP-strip (Fig.5(a)). As discussed in the previous section, it corresponds to the screening currents in horizontal branches of the grid, decreasing towards the interior of the circuit.

With increasing $H_z$, the step-wise changes in the $B_z(y)$ scans appear across the W-grid (Fig. 5(b-c)) due to the flux channels extending from the corners of contact pads. They are accompanied by periodic ~1-2 G oscillations with minima above the S-square projections (marked by arrows in Fig.5(b)) and maxima between them. In large fields, the oscillations become regular on a smooth background which decreases towards the GP-strip (Fig.5(d)). Here the W-grid starts acting as a quasi-homogeneous screening medium.

With decreasing $H_z$ (Fig. 5(e)), the field around the contact pad drops below $H_z$, the oscillations due to the W-grid remain, but their smooth background curve starts decreasing towards the pads.

The oscillations due to the S-squares remain, although with some asymmetry due to the inversion of the direction of currents in the W-grid wires during field reduction. In zero field (Fig. 5(f)), periodic oscillations over the grid become very small and the background curve shifts below zero.

*d) Flux distributions in the multilayered contact pads*

The periodic flux patterns are observed also in the contact pads, where continuous Nb film rectangles in layers M0, M2, and M7 alternate with rectangles of orthogonally oriented strips in intermediate layers M1, M3-M6, as shown in Fig.7. The x,y-positions of the pads in different layers coincide, and they are all connected by Nb vias. The z-projections of the SC strips from intermediate layers leave open square regions (Sq-regions), which are covered by Nb film only in the continuous pads.

Figure 8 illustrates the flux penetration in this complex multilayer structure. At low fields, the whole pad area remains flux-free showing the dominating screening effect of the layers with continuous Nb rectangles (Fig. 8(a)). However, with increasing $H_z$, the flux enters the pad area, occupying the Sq-regions between the strip projections; see Fig.8(b). The flux inside the Sq-regions concentrates at the side of the vertical strips running between the short edges of the pad (see small arrows in panel (b)) due to the screening by the strips in the layer M5, which maintain a flux-free space deeper within the pad. Similarly, horizontal strips of other intermediate layers restrict the flux spreading outside of the Sq-regions in the y-direction. Although the strips are located in different layers of the pad, they are all connected by superconducting interlayer vias, and the picture of flux distribution in the Sq-regions is reminiscent of that in square holes near the edge of a continuous SC sample. In general, within the pad the penetrating flux lines are bending around the horizontal and vertical Nb strips in different layers and yield the concentration of vortices in the continuous Nb rectangles above the Sq-regions.

With further increasing $H_z$ (Fig. 8(c-d)), the flux penetration into the Sq-regions progresses inside the pad and the flux density in them increases. Gradually, this periodic square pattern occupies the entire pad area, Fig. 8(e). Interestingly, the overall front of the flux penetration in Fig. 8(c-e) follows the pillow-shaped pattern typical for homogeneous rectangular SC samples [21].

After a subsequent reduction of $H_z$, the flux decreases in the Sq-regions, while the maximum $B_z$ remains trapped in the network of strip projections (Fig. 8(f)). Now the flux is concentrated at the sides of the strips facing the center of the pad corresponding to the inversion of the current direction along the strips. Again, the average flux distribution resembles the critical state pattern in a continuous rectangle.

*e) Field cooling in small magnetic fields*

Finally, we address the distribution of the trapped flux in the circuit after cooling in a small perpendicular field. To resolve the anticipated tiny effects for the SC sample cooled in $H_z$~1 Oe, we accumulated images with a hundred of exposures and referenced them by images taken at $T>T_c$ in the same field. Figure 9 shows strongly contrasted FC pictures of the GP-strips and contact pad regions. Clearly, magnetic flux is partially expelled from the major GP-strip areas into the slits (see brighter contrast along the slit-lines in Fig. 9(a)), confirming their intended flux trapping action. At the same time, the flux is removed more efficiently near the edges of the GP-strips; see the darker contrast regions pointed by arrows in Fig. 9(a).

In the contact pads, the flux is expelled stronger from the areas of the Nb strips located in different layers and is accumulated in the Sq-regions between them indicated by curved arrows Fig. 9(b). Here the field is enhanced due to the flux expulsion from the strips. Due to the weak signal, it is difficult to give a quantitative description of the observed flux changes in the FC patterns observed in different parts of the circuit.

## Conclusions

In this work, we present the first successful attempt to image flux patterns in a real superconductor digital integrated circuit consisting of eight patterned superconducting layers with a complex distribution of components of different sizes and shapes. Although our technique maps magnetic induction above the top surface of the sample, the observed $B(r)$- distributions reveal the effects of flux screening and trapping not only in the upper superconducting layer but also of the "hidden" lower layers in the complicated 3D structure.

We have analyzed the role of different components which act cooperatively and induce elaborate meandering flux flow around separate design elements and concentrate magnetic flux in the gaps between them. The resulting inhomogeneous flux distributions create an intricate vortex pattern in large ground planes. This pattern then affects complex maps of the trapped flux in the decreasing fields. Overall, the averaged magnetization distributions observed in the large electrically connected regions of the circuit resemble the flux entry and exit in inhomogeneous but continuous slices of a "superconducting Swiss cheese".

Our observations have shown that the square grid of superconducting wires surrounding the main circuit area provides an efficient flux screening in relatively small magnetic fields typical for the circuit operation environment. Concurrently, the slit-shaped moats in the ground planes introduce a large enhancement of the local induction and can be a source of secondary vortices even in small external fields. Also, the in-line arrangement of closely spaced slits is prone to instabilities in the flux penetration. At the same time, such slits are needed to decrease the effective width of extended ground planes, reducing the flux concentration at their edges and assisting the removal of residual Abrikosov vortices in the field-cooled cases. The obtained results can provide guidance in the careful redesign of the shape and relative position of different elements in the next generation of multilayered superconducting devices.

## Methods

The shift register circuit for this study was designed and manufactured at MIT Lincoln Lab as described in [8]. The 5x5 mm chip containing six registers was mounted on a cold finger of the Montana optical cryostat, covered by a magneto-optical garnet indicator film with a large Verdet constant, and imaged in a polarized light microscope. Using home-made software and a high sensitivity camera, PIXIS 1024, we recorded images of magnetic flux distributions in changing magnetic fields or after field cooling, which were then transformed into induction maps, $B_z(r)$, using image intensity-$B_z$ calibration.

## Acknowledgements

We are thankful to Dr. Rory Perkins for his interest in and support of this work. We are also grateful to Prof. Vasili Semenov for the original design of the shift registers, and to Drs. Ravi Rastogi and David Kim for overseeing the circuits fabrication runs.

The research by TR, AG, BJ, and VVV was sponsored by the Army Research Office under Grant No. W911NF-24-1-0145. The work at MIT Lincoln Laboratory was supported under Air Force Contract No. FA8702-15-D-0001 or FA8702-25-D-B002. The views and conclusions contained in this document are those of the authors and should not be interpreted as representing the official policies, either expressed or implied, of the Army Research Office and the Air Force or the U.S. Government. Delivered to the U.S. Government with Unlimited Rights, as defined in DFARS Part 252.227-7013 or 7014 (Feb 2014). Notwithstanding any copyright notice, U.S. Government rights in this work are defined by DFARS 252.227-7013 or DFARS 252.227-7014 as detailed above. Use of this work other than as specifically authorized by the U.S. Government may violate any copyrights that exist in this work. The U.S. Government is authorized to reproduce and distribute reprints for Government purposes notwithstanding any copyright notation herein.

## Author contributions statement

TR- major imaging experiments and discussion of results; AG- recovery of the sample structure from gds files and discussion of results; BJ-analysis and discussion of results, writing the paper; JM -circuit design discussion of results; ST -circuit design, discussion of results and writing the paper; VVV – magneto-optical imaging, analysis of results and writing the paper.

## Additional information

**Competing interests:** The authors declare no competing interests.

**Data availability**: The data that support the findings of this study are available upon request from the authors.

## Figures & captions

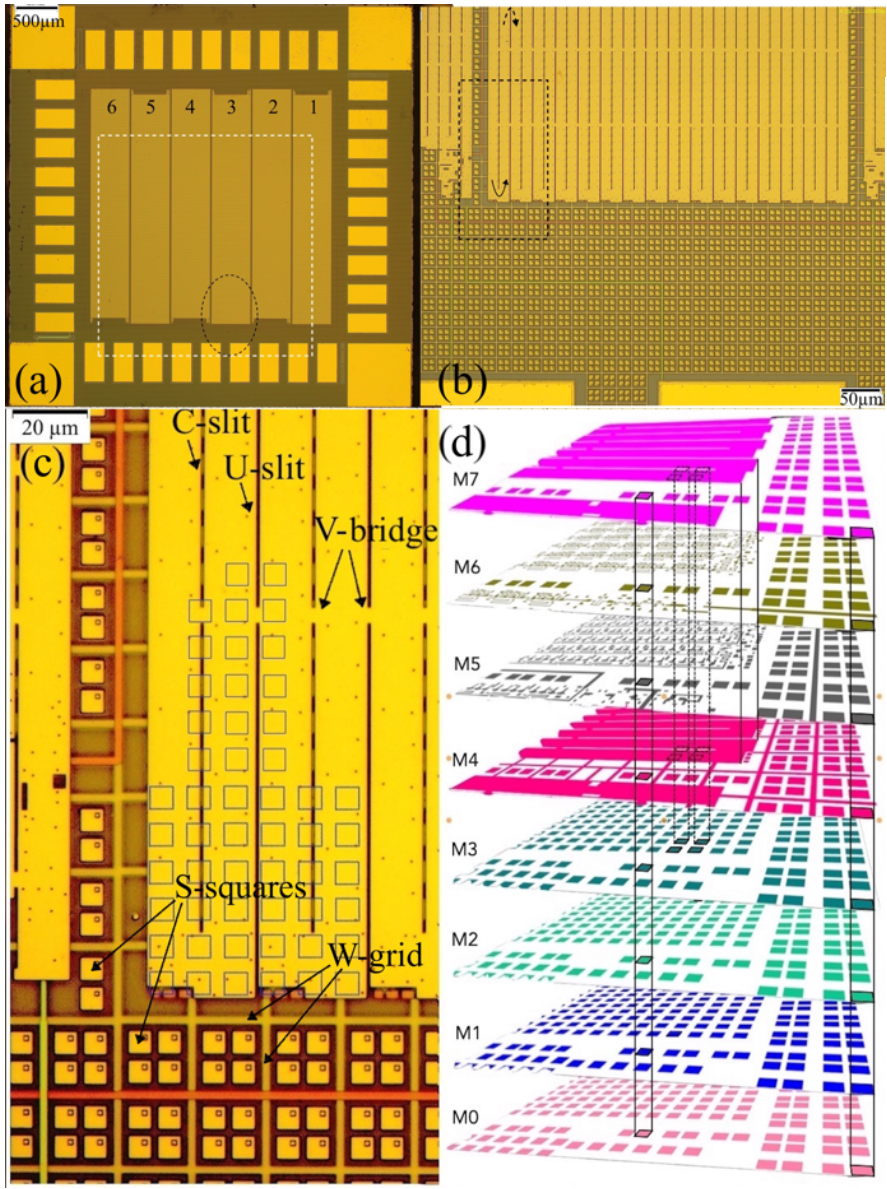

**Fig. 1.** Optical images of the circuit and its components: (a) picture of the circuit surface; (b) magnified image of the area outlined by a dashed oval in (a); (c) enlarged picture of the region outlined by the dashed rectangle in (b); (d) 3D reconstruction of the circuit design in the region shown in (b).

In (a), bright rectangles at the periphery are contact pads. Ground plane strips (GP-strips) in the middle, marked 1-to-6, are perforated with slits of different length and spacing as described in the text. The wire grid (W-grid in (c)) surro- nds the GP-strips; each grid cell contains small Nb squares (S-squares in (c)) residing in lower layers and functioning as a metal fill required for the circuit fabrication. Slit lines intermittently reach the GP-strip edges (U-slits in (c)) or terminate at some distance (C-slits), which changes on the opposite strip edge as sketched by bent arrows at the top and bottom of panel (b). The white dashed square in (a) marks the area of magneto-optical images shown in Fig. 2. Panel (d) shows the location of the circuit elements in different layers. Homogeneous colors correspond to the regions of continuous 200-nm-thick Nb film. The main digital electronics components are located in layers M5 and M6 (see description in the text) between the GP-strips in layers M4 and M7. All layers contain 6 μm x 6 μm Nb S-squares outside of the GP-strip regions and also under the GP-strips in four bottom layers M0-M3. Vertical lines in (d) show the z-projections of the S-squares. These projections cover some slits in the GP-strips (C-slits), as shown by blue squares in (c), and leave their neighbors uncovered (U-slits).

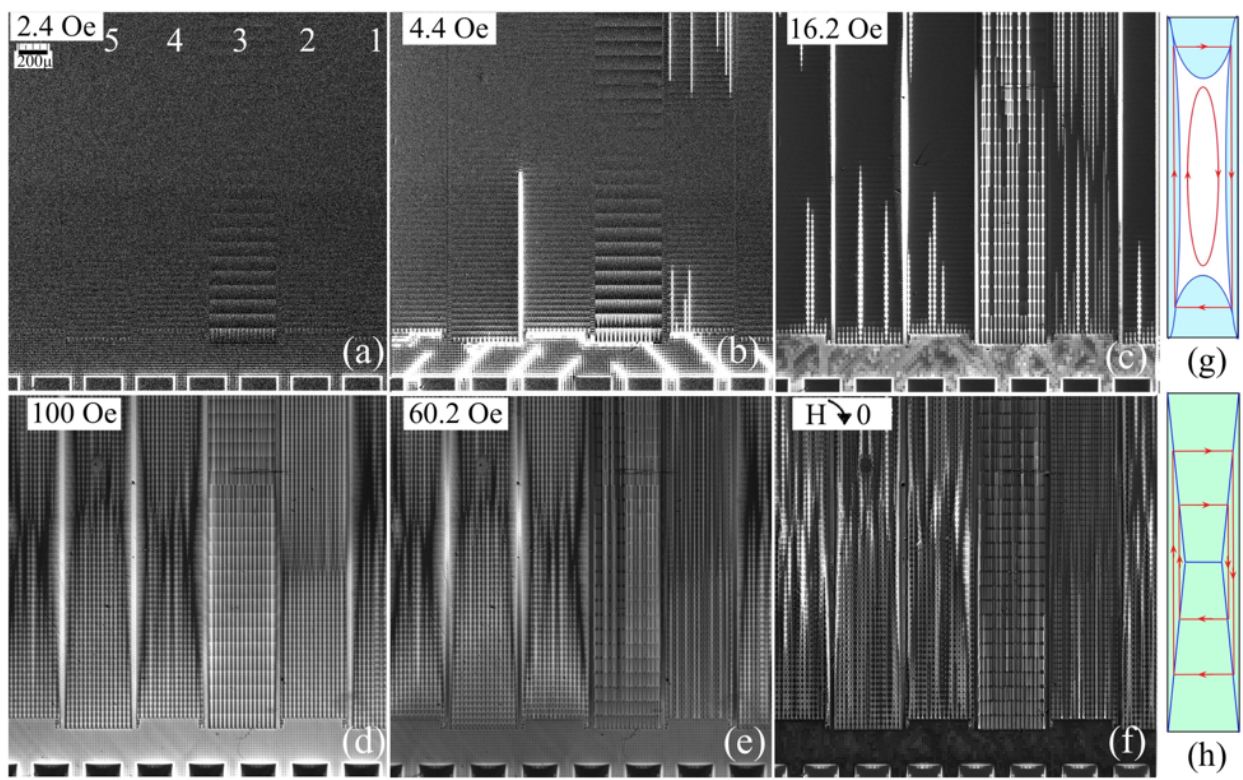

**Fig. 2.** Magnetic flux patterns in the area marked by the dashed square in Fig. 1(a) in increasing (a-d) and decreasing (e-f) magnetic fields at T=5 K. Values of $H_z$ are shown in the panels. The $B_z$-scale (range of image intensity) in successive panels is changed for clear visualization of the flux inhomogeneities; $B_z$ values are discussed in section c). Panels (g) and (h) sketch the overall flux (blue regions) and current (red lines) distributions in the GP-strips, resembling those observed at magnetization of anisotropic SC rectangles.

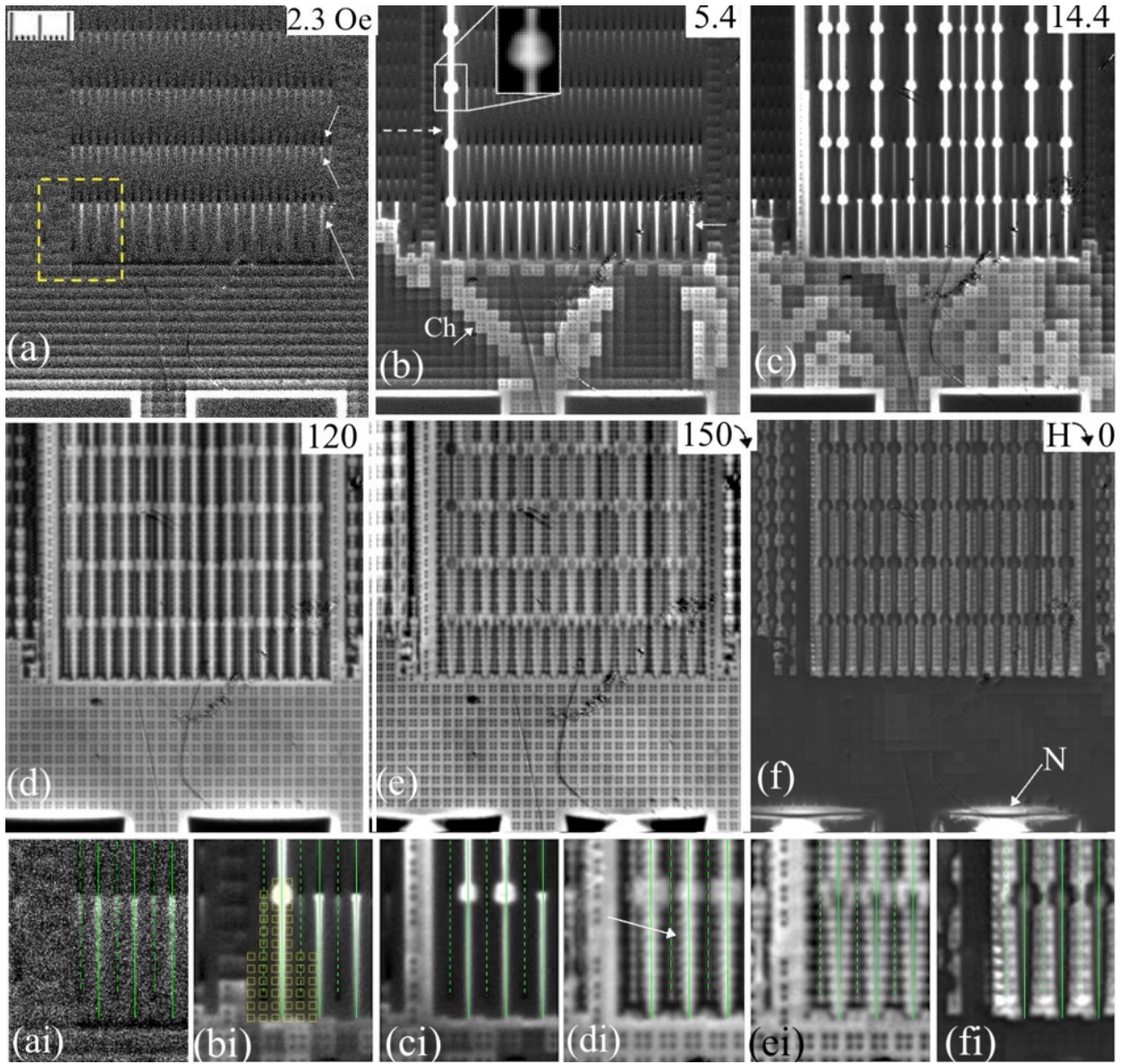

**Fig. 3.** Magnetic flux patterns near the edge of GP-strip #3 in the increasing (a-d) and decreasing (e-f) fields at 7.5 K. Values of $H_z$ are shown in the panels. Enlarged fragments of (a-f), in the area outlined by a yellow rectangle in (a), are shown at the bottom with appropriate indices (ai-fi). Positions of the U- and C-lines are marked in (ai-fi) by solid and broken green lines. Panel (bi) also shows locations of some of the S-squares in the lower layers. Up-arrows in (a) point to bright regions of enhanced $B_z$ formed below the V-bridges in the slit lines, while Down-arrow points to the dark region of negative $B_z$ above the V-bridges. Dashed arrow in (b) shows the abruptly filled line of U-slits and the arrow with Ch points to the flux channel in the W-grid extended at $45^0$ from the contact pad corner. The arrow with N in panel (f) points to the region of negative $B_z$ around the contact pad carrying large positive trapped flux. In panel (di), the arrow points to the toothy structure extending from the U-slits in large fields.

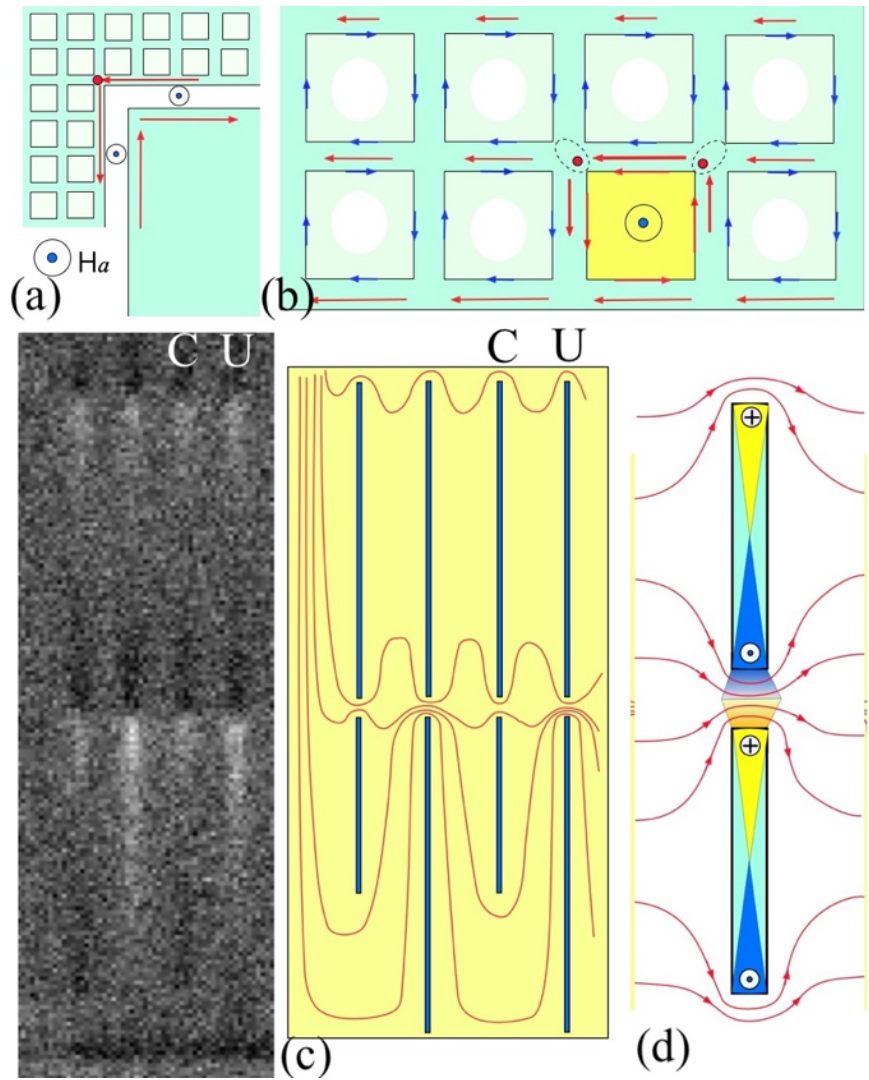

**Fig. 4.** Sketches of the SC screening currents: (a) around the contact pads; (b) in the wire grid; (c) near the C- and U-slits in the GP-strips. Panel (b) also shows separately the current component screening the square apertures in the W-grid (blue arrows) and the average currents in the wires (red arrows). The length and thickness of red arrows in (a-b) correspond to the current strength which decreases towards the interior of the sample. The current crowding regions are marked by red dots. The yellow square in (b) illustrates a cell in the W-grid with entered flux and, correspondingly, inverted screening currents around it. Panel (c) illustrates the increased current density in V-bridges of U-slits, which promotes the easier flux transfer along the U-lines extended to the GP-strip edge. Panel (d) shows the sketch of appropriate fields in the slits, corresponding to the observed $B_z$-map shown on the left of (c).

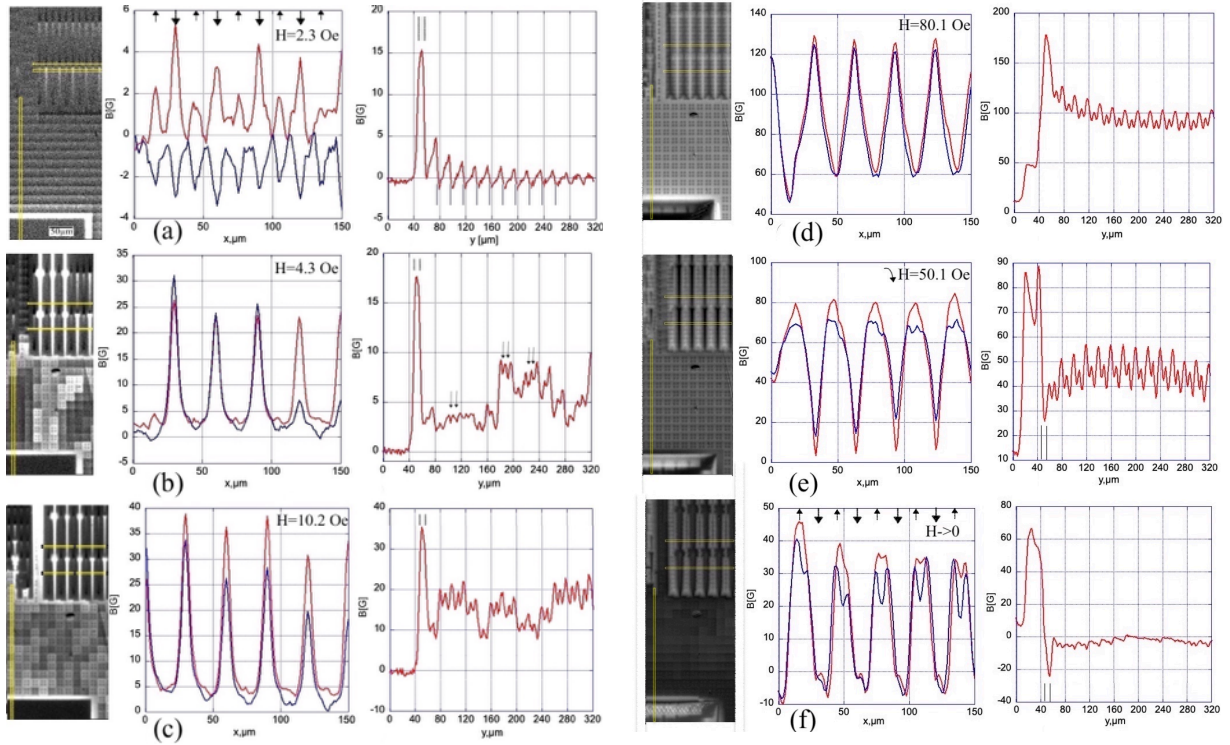

**Fig. 5.** $B_z$-profiles in the increasing (a-d) and decreasing (e-f) fields at T=7.5 K . Left B(x) plots in each panel are $B_z$-profiles across the slits in GP-strip #3. Right B(y) plots in each panel are $B_z$-profiles along the wire grid. The positions of scans are shown by yellow lines in the images on the left. Field values are shown in the panels. $B_z$ plots across the slits are measured below the V-gaps (red curves) and above the gaps (blue). Scans along the W-grid start in the contact pad area. Position of U-lines and C-lines are marked in the B(x) plot of panel (a) by ↓ and ↑ arrows, respectively. Y-position of the band around the contact pad and locations of the W-grid wires are marked by vertical lines above and below the B(y) line of panel (a), respectively. Small arrows in the B(y) plot of panel (b) mark the location of S-squares.

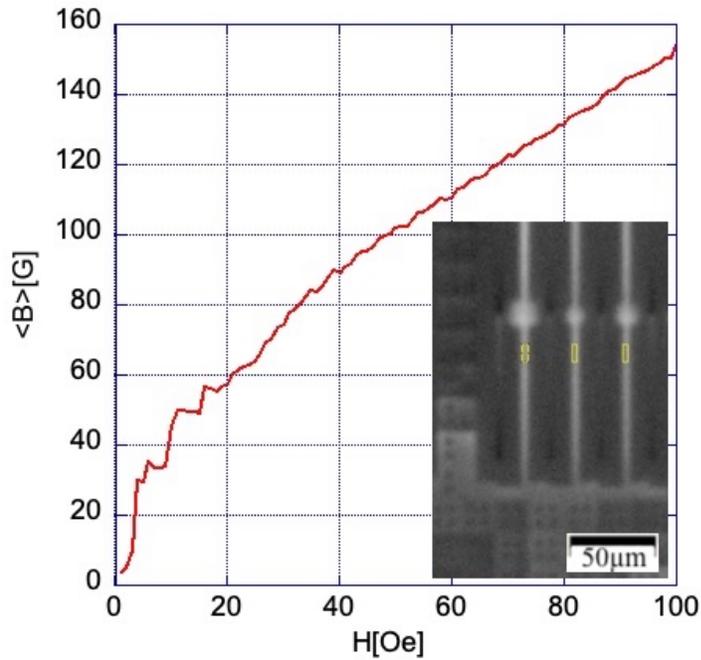

**Fig. 6.** The average flux density, $<B_z>$, in U-slits as a function of the applied field. The measurement spots in three slits near the edge of the GP-strip are shown in the inset.

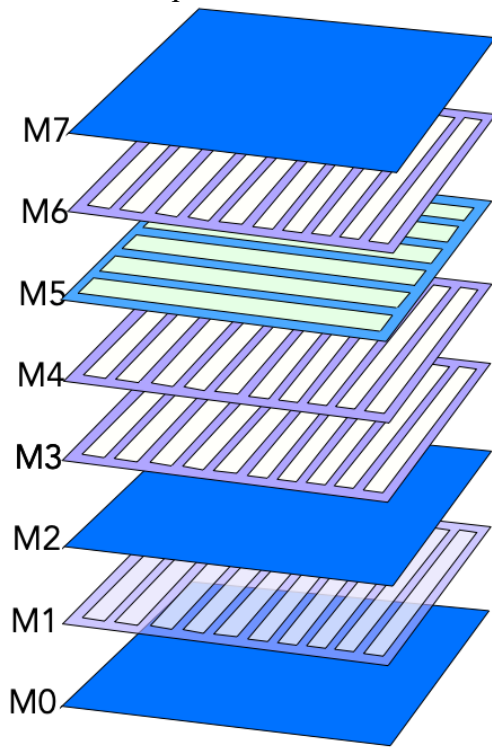

**Fig.7.** Design of the contact pads. The pad rectangles in different layers are composed of the continuous Nb film (layers M0, M2, and M7) and Nb strips, which are parallel (layer M5) or perpendicular (in the remaining layers) to the long side of the rectangle. In the circuit, the pad features in different layers are aligned on top of each other and connected by superconducting vias.

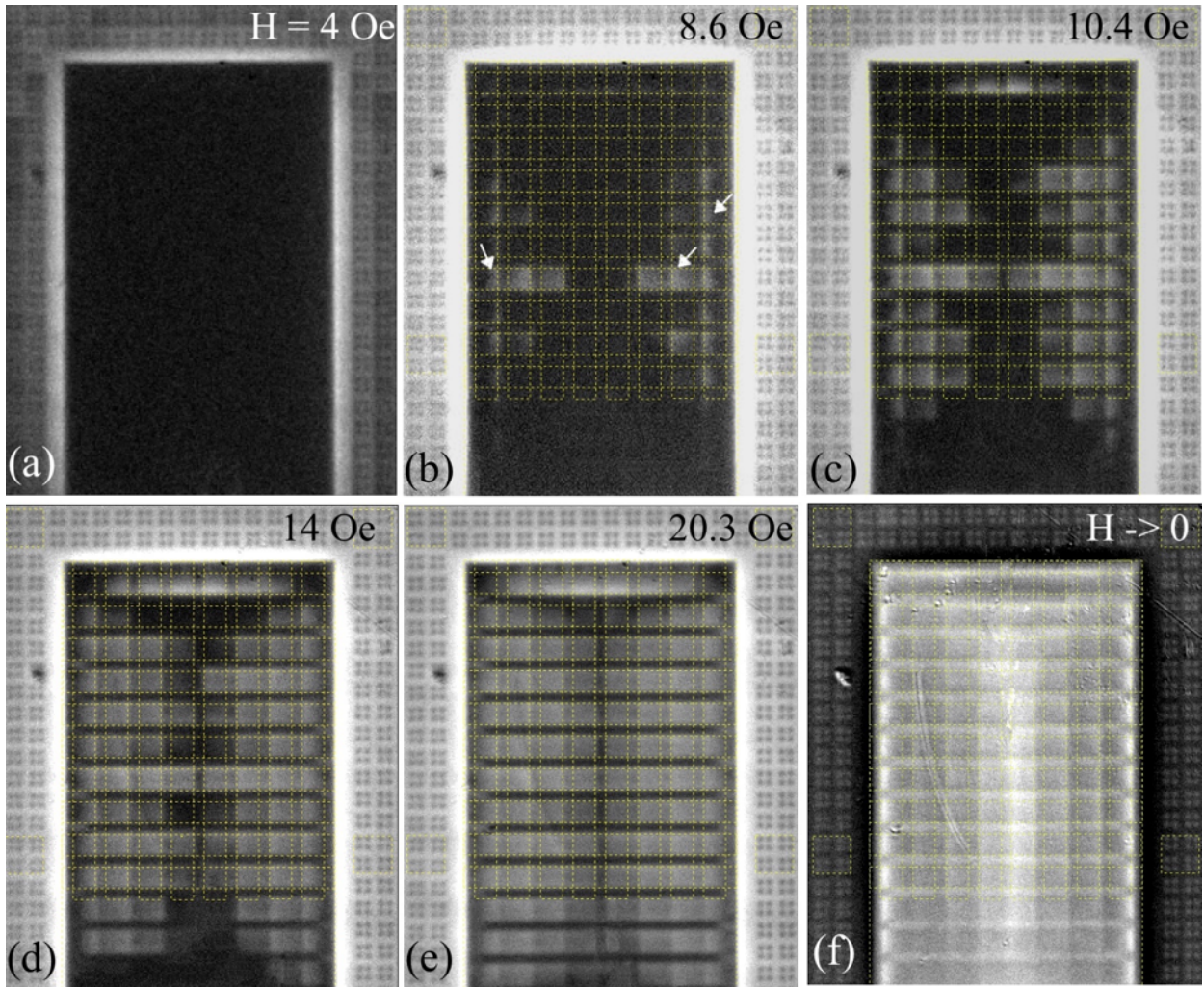

**Fig. 8.** Flux distributions in the contact pad in the increasing field (a-e) and after switching the field off at T=7.5 K. Dotted lines in the pictures show the projections of Nb strips in different layers of the pads. Square regions in the dash-line lattice correspond to an "empty" space between the strips, which is covered by continuous Nb rectangles in layers M7, M2, and M0. Arrows in panel (b) point to the enhanced $B_z$ at the vertical projections of the strips. In (b)-(e) the penetrating flux is concentrated in the square regions between the strips. From the same regions, the flux preferentially exits at decreasing field, as shown in (f). In the increasing and decreasing fields, the average flux distribution acquires a pillow-shape, characteristic for SC rectangles in the critical state.

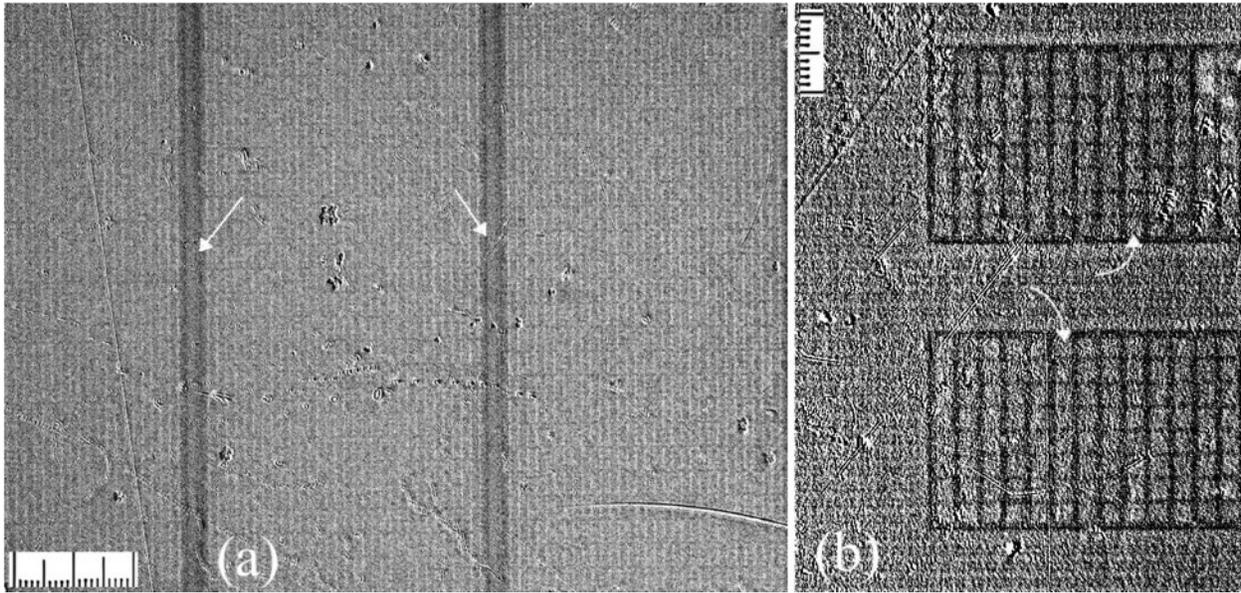

**Fig. 9.** Flux distributions in the GP-strips (a) and in the contact pads (b) after field cooling from 9.5 to 5 K in $H_z=1$ Oe. Although the field is not switched off, the flux is partially expelled from superconducting regions. Arrows in (a) show dark regions of the preferential flux exit near the GP-strip edges. In (b) the major flux expulsion is along the projections of Nb strips in the pad layers. In turn, the maximum trapped flux remains between the strips, as pointed by curved arrows in (b).